\documentclass[%
 aip,
 jmp,%
 amsmath,amssymb,
 reprint,%
]{revtex4-1}

\usepackage{graphicx}
\usepackage{dcolumn}
\usepackage{bm}

\begin{document}

\preprint{AIP/123-QED}

\title[]{Quantum Hydrodynamics Approach to The Research of Quantum Effects and Vorticity Evolution in Spin Quantum Plasmas}

\author{Trukhanova M. Iv.}
 \altaffiliation[ ]{M.V.Lomonosov Moscow State University, Faculty of Physics, Leninskie Gory,  Moscow, Russia}
\email{mar-tiv@yandex.ru}

\date{02.01.2013}

\begin{abstract}
In this paper, we explain a magneto quantum hydrodynamics (MQHD) method for the study of
the quantum evolution of a system of spinning fermions in an external electromagnetic field.
The fundamental equations of microscopic quantum hydrodynamics (the momentum balance
equation, the energy evolution equation and the magnetic moment density equation) are derived from the many-particle microscopic
Schredinger equation with a spin–spin and Coulomb modified Hamiltonian. Using the
developed approach, an extended vorticity evolution equation for the quantum spinning plasma
is derived. The effects of the new spin forces and spin–spin interaction contributions on the
motion of fermions, the evolution of the magnetic moment density, and vorticity generation
are predicted. The influence of the intrinsic spin of electrons on whistler mode turbulence is
investigated. The results can be used for theoretical studies of spinning many-particle systems,
especially dense quantum plasmas in compact astrophysical objects, plasmas in semiconductors,
and micro-mechanical systems, in quantum X-ray free-electron lasers.  \end{abstract}

\pacs{52.35.We, 67.10.-j}
\maketitle
\section{Introduction}
The spinning quantum fluid plasma is becoming of increasing
current interest \cite{1} - \cite{7}.
Hydrodynamics  equations of a spinning fluid for the Pauli equation with the quantum particle angular
momentum spin was presented since the pioneering works by Takabayashi and Vigier   \cite{8} - \cite{11}. The vector representation of non-relativistic spinning particle leads to appearance of new quantum effects had been separated as non-linear terms which arises from the inhomogeneity of spin distribution.  The extension of the interpretation to developed approach had been carried in the \cite{110} - \cite{1100}.

The quantum effects in plasma can be represented by three main
quantum corrections. The first is a quantum force, the multiparticle quantum Bohm or Madelung potential,  proportional to
powers of $\hbar$ and produced
by density fluctuations \cite{111}, \cite{112}.  The second is associated with the quantum particle
angular momentum spin by the possible inhomogeneity of the external and spin magnetic fields. In
the momentum balance equation this force appears through the magnetization energy \cite{1}. And the latter force   associated only with the spin magnetic moment of the particle \cite{8}.

The most interesting and
defining features of a quantum spinning plasmas can be derived from the vorticity equation.   It had been derived by \cite{12} that the vorticity,  constructed from spin field of a quantum spinning plasma, combines
with the classical generalized vorticity  to yield
a new grand generalized vorticity that obeys the standard vortex dynamics.

Astrophysics is also a rapidly growing field of research. It is important that the consequences of turbulent
plasma movement in the solar photosphere lead to the generation of vorticity, while magnetic
vortices are produced by magnetic tension. For example, magnetohydrodynamics (MHD) simulations
of magnetoconvection have been used to analyze the generation of small-scale vortex motions
in the solar photosphere. Using the vorticity equation, combined with G-band radiative diagnostics,
it has been shown that two different types of photospheric vorticity, magnetic and non-magnetic,
are generated in the domain \cite{13}. The presence of vortex motions for the astrophysics had been developed in \cite{14} - \cite{15}.

The extracting coherent vortices
out of turbulent flows  had been applied to simulations of resistive drift-wave turbulence in magnetized plasma \cite{17}. The quasi-hydrodynamic and quasi-adiabatic regimes had been investigated.

The formation and dynamics of dark solitons and vortices in quantum electron plasmas had been studied in \cite{18}.   A pair of equations comprising the nonlinear Schrödinger and Poisson system of equations, which conserves the number of electrons as well as their momentum and energy had been used.   It had been shown that the gradient "free-energy" contained in    equilibrium spin vorticity can cause electromagnetic modes, in particular the light wave \cite{19}.

The collective electron angular momentum spin effects in spinning quantum plasmas can be investigated
using insights from quantum kinetic theory or some effective theory. We propose a method
of quantum hydrodynamics that allows one to obtain a description of the collective effects in magnetized
quantum plasmas in terms of functions in physical space. The fermion model was developed in
Refs. \cite{1}, \cite{2}, \cite{20} and \cite{21}. The waves in the magnetized plasma with the spin had been studied in \cite{20} exploring of new quantum hydrodynamics method of the generation wave in the plasma. The new formalism given in this references  had been used in this article for studying of vorticity evolution in the magnetized plasma with the spin.
A quantum mechanics description for systems of N interacting spinning particles is based upon
the many-particle Schrödinger equation (MPSE) that specifies a wave function in a 3N-dimensional
configuration space. As wave processes, processes of information transfer, and other spin transport
processes occur in 3D physical space, it becomes necessary to turn to a mathematical method of
physically observable values that are determined in a 3D physical space. To do this, we should derive
the fundamental equations that determine the dynamics of functions of three variables, starting from
MPSE. This problem has been solved with the creation of a many-particle quantum hydrodynamics
(MPQHD) method.

In this article for studying of vorticity and spin vortex effects we  generalize and use the method of the
many-particle quantum hydrodynamics  MQHD approach.  We derive the fundamental balance equation, the magnetic moment evolution equation and new  vorticity dynamics equation and the magnetic vortex evolution equation for the magnetized quantum plasmas.

\section{\label{sec:level1}Fundamental equations of the fermion quantum hydrodynamics}
  In this section we derive the system of magneto quantum hydrodynamics (MQHD) equations for charged and neutral particles from the many-particle microscopic Schr¨odinger equation

  \begin{equation}
            i\hbar\frac{\partial\psi_s(R,t)}{\partial t}=(\hat{H}\psi)_s(R,t),
  \end{equation}
where $R=(\vec{r}_{1},...,\vec{r}_{N})$. We consider a system of N interacting fermions with equal masses $m_j$, charged and proper magnetic moments in an external electromagnetic field. A state of the system of $N$ fermions is determined by a wave function in the 3N-dimensional configuration space, which is a $rank-N spinor$
            \begin{equation}
            \psi_s(R,t)=\psi_{s_1,s_2,....s_N}(\vec{r}_{1},...,\vec{r}_{N},t).
  \end{equation}
              The Hamiltonian has the form
              \begin{equation} \label{H}    \hat{H}=\sum^{N}_{j=1}({\frac{\hat{D}^{2}_{j}}{2m_j}+q_j\varphi_{j,ext}-\mu_j\hat{\sigma}^{\alpha}_{j}B^{\alpha}_{j,ext}})+
\end{equation}
\[
\qquad\qquad\qquad+\frac{1}{2}\sum^{N}_{j\neq k} q_jq_kG_{jk}-\frac{1}{2}\sum^{N}_{j\neq k,k}\mu^{2}_{j}F^{\alpha\beta}_{jk}\hat{\sigma}^{\alpha}_{j}\hat{\sigma}^{\beta}_{k},
\]
\[
\]
where $\mu_j=g\mu_B/2$, $\mu_{jB}$ - is the electron or positron magnetic moment  and $\mu_{jB}=q_j\hbar/2m_jc$ - is the Bohr magneton,  $q_j$ stands for the charge of electrons $q_e=-e$ or for the charge of positrons $q_p=e$, and $\hbar$ -is
the Planck constant, $g\simeq 2.0023193$. The covariant derivative operator
is

      \begin{equation} \label{D}    \hat{D}^{\alpha}_j=-i\hbar\hat{\nabla}^{\alpha}_j-\frac{q_j}{c}A^{\alpha}_j,\end{equation}
where     $\vec{A}_{ext}, \varphi_{j,ext} $  -  are   the vector and scalar potentials of
external electromagnetic field.

Green's functions of  the $Coulomb$ and $Spin - Spin$ interaction  are\begin{equation}  G_{jk}=\frac{1}{r_{jk}}, \qquad  F^{\alpha\beta}_{jk}=4\pi\delta_{\alpha\beta}\delta(\vec{r}_{jk})+\partial^{\alpha}_{j}\partial^{\beta}_{k}\frac{1}{r_{jk}}.
                                \end{equation}

The first step in the        construction of MQHD apparatus is
to determine the concentration of particles in the neighborhood
of $\vec{r}$ in a physical space. If we define the concentration
of particles as quantum average of the concentration
operator in the coordinate representation $\hat{\rho}=\sum_j\delta(\vec{r}-\vec{r}_j)$ we obtain

                               \begin{equation}
\rho(\vec{r},t)=\sum_{S}\int dR\sum^{N}_{j}\delta(\vec{r}-\vec{r}_{j})\psi^{+}_s(R,t)\psi_s(R,t),
\end{equation}

                                 Differentiation of $\rho(\vec{r}, t)$ with respect to time and applying
of the Schr¨odinger equation with Hamiltonian \ref{H} leads
to continuity equation

                        \begin{equation} \label{n}
\frac{\partial \rho(\vec{r},t)}{\partial t} +\vec{\nabla}\vec{\jmath}(\vec{r},t)=0, \end{equation}
                        where the current density takes a form of

 \begin{equation} \label{J}
\jmath^{\alpha}(\vec{r},t)=\sum_{S}\int dR\sum^{N}_{j}\delta(\vec{r}-\vec{r}_{j})\frac{1}{2m_{j}}(\hat{D}^{+\alpha}_{j}\psi^{+}_s(R,t)\psi_s(R,t) \end{equation}
                                         \[
                                        \qquad\qquad\qquad\qquad\qquad +\psi^{+}_s(R,t)\hat{D}^{\alpha}_{j}\psi_s(R,t)),
                                         \]

                                         A momentum balance equation can be derived by differentiating
current density \ref{J} with respect to time
                                           \begin{equation} \label{jjj}
\partial_t\jmath^\alpha(\vec{r},t)+\frac{1}{m}\partial_{\beta}\Re^{\alpha\beta}(\vec{r},t)=\frac{q}{m}\rho(\vec{r},t)E^{\alpha}_{ext}(\vec{r},t)\end{equation}
\[+\frac{q}{mc}\varepsilon^{\alpha\beta\gamma}j_{\beta}(\vec{r},t)B^{\gamma}_{ext}(\vec{r},t)-\frac{1}{m}\int d\vec{r}^{'}q^2\partial^{\alpha}G(\vec{r},\vec{r}^{'})\rho_2(\vec{r},\vec{r}^{'},t)
\]
\[\qquad +\frac{1}{m}M_{\beta}(\vec{r},t)\partial^{\alpha}B^{\beta}_{ext}(\vec{r},t)
+\frac{1}{m}\int d\vec{r}^{'}\partial^{\alpha}F^{\gamma\delta}(\vec{r},\vec{r}^{'})M^{\gamma\delta}(\vec{r},\vec{r}^{'},t).\]

 \begin{equation}  \label{P}
\Re^{\alpha\beta}(\vec{r},t)=\sum_{S}\int dR\sum^{N}_{j=1}\delta(\vec{r}-\vec{r}_{j})\frac{1}{4m_{j}}(\psi^{+}(R,t)\hat{D}^{\alpha}_{j}{D}^{\beta}_{j}\psi(R,t)
\end{equation}
\[\qquad\qquad\qquad+(\hat{D}^{\alpha}_{j}\psi(R,t))^{+}D^{\beta}_{j}\psi(R,t)+h.c.)\]
represents the momentum current density tensor.

Momentum balance equation \ref{jjj} contains the particle magnetic moment density    \cite{1}
\begin{equation}
M^{\alpha}(\vec{r},t)=\sum_{s}\int dR\sum^{N}_{j=1}\delta(\vec{r}-\vec{r}_{j})\mu_j\psi^{+}_s\hat{\sigma}^{\alpha}_j\psi_s,
\end{equation}

The $Coulomb$ and $Spin-Spin$ interactions between the particles are represented in Eq. \ref{jjj} by the terms where
 \begin{equation}  \label{F}
    \rho_2(\vec{r},\vec{r}^{'},t)=\sum_s\int dR\sum^{N}_{j\neq k} \delta(\vec{r}-\vec{r}_{j})\delta(\vec{r}^{'}-\vec{r}_{k})\psi^{+}(R,t)\psi(R,t), \end{equation}
is the two-particle probability density for the occurrence of two particle in the neighborhoods of the points $\vec{r}$ and $\vec{r}^{'}$  normalized by $N(N-1)$, and two-particle tensor of the magnetic moment density
  \begin{equation}  \label{F2}
   M^{\alpha\beta}(\vec{r},\vec{r}^{'},t)=\sum_s\int dR\sum^{N}_{j\neq k} \delta(\vec{r}-\vec{r}_{j})\delta(\vec{r}^{'}-\vec{r}_{k})\end{equation}
                                         \[         \qquad\qquad\qquad\qquad\qquad\qquad \times\mu_j\mu_{k}\psi^{+}\hat{\sigma}^{\alpha}_{j}\hat{\sigma}^{\beta}_{k}\psi(R,t). \]
     Differentiation of $M^{\alpha}$ with respect to time and applying
of the Schr¨odinger equation with Hamiltonian \ref{H} leads
to magnetization equation.
The equation representing the non-relativistic evolution
of $spin-1/2$ motion takes a form of

              \begin{equation}  \label{magnit}
\partial_t M^{\alpha}(\vec{r},t)+\partial_{\beta}\Im^{\alpha\beta}_{M}=\frac{2\mu}{\hbar}\epsilon^{\alpha\beta\gamma}M^{\beta}(\vec{r},t)B^{\gamma}_{ext}(\vec{r},t)
\end{equation}
           \[   \qquad\qquad+\frac{2\mu}{\hbar}\epsilon^{\alpha\beta\gamma}\int d\vec{r}^{'}F^{\gamma\delta}(\vec{r},\vec{r}^{'})M^{\beta\delta}(\vec{r},\vec{r}^{'},t)
            \]
           where the tensor of the magnetic moment flux density is

           \begin{equation}  \label{JM}
\Im^{\alpha\beta}_{M}(\vec{r},t)=\sum_{S}\int dR\sum^{N}_{j=1}\delta(\vec{r}-\vec{r}_{j})\frac{\mu_j}{4m_{j}}(\psi^{+}\hat{\sigma}_j^{\alpha}\hat{D}^{\beta}_{j}\psi+
\end{equation}
\[\qquad\qquad\qquad\qquad\qquad+(\hat{\sigma}_j^{\alpha}D^{\beta}_{j}\psi)^{+}\psi)(R,t).\]

The spinning quantum magnetohydrodynamics should explain the vorticity evolution. The main
idea of this paper was to create a hydrodynamics foundation for the vortex dynamic in the context
of spinning quantum plasma. We use the MPQHD approach to receive the equations for the particle
vorticity density, obeying the standard vortex dynamics. We  determine the vorticity density vector of particles in the neighborhood
of $\vec{r}$ in a physical space  as

                          \begin{equation} \label{J2}
\Omega^{\alpha}(\vec{r},t)=\sum_{S}\int dR\sum^{N}_{j}\delta(\vec{r}-\vec{r}_{j})\frac{\varepsilon^{\alpha\beta\gamma}}{2m_{j}}\hat{\nabla}_{j\beta}(\hat{D}^{+\gamma}_{j}\psi^{+}_s\psi_s \end{equation}
                                         \[
                                        \qquad\qquad\qquad\qquad\qquad \qquad\qquad +\psi^{+}_s\hat{D}^{\gamma}_{j}\psi_s)(R,t),
                                         \]
where we construct the vorticity density   in term of the wave function, we denote the macroscopic vorticity density as $\vec{\Omega}=\vec{\nabla}\times\vec{j}$, as will be shown below.  The classical generalized vorticity density  $\vec{\Omega}$ can be defined as the curl of  the current density.    But in article \cite{12} the ordinary vorticity of the plasma  is proportional to the curl of the  flow velocity of the fermions  (vorticity
have the dimensions of the magnetic field).On the other hand vorticity can be defined as the curl of velocity $\vec{\omega}=\vec{\nabla}\times\vec{\upsilon}$ \cite{13}. Our idea is that we use the definition \ref{J2} and MQHD method  based upon the many-particle Schr¨odinger equation to derive the generalized dynamical equation for classical  vorticity $\vec{\omega}=\vec{\nabla}\times\vec{\upsilon}$  similar to \cite{13}  and \cite{14} (see below), but contained the information about interactions inside the fluid.

 \subsection{\label{sec:level2}Velocity field}

The velocity of $j-th$ particle $\vec{\upsilon}_j$ is determined by
equation

\begin{equation} \label{z4}
         \vec{\upsilon}_j=\frac{1}{m_j}(\vec{\nabla}_jS-i\hbar\varphi^+\vec{\nabla}_j\varphi)-\frac{q_j}{m_j c}\vec{A}_j,
           \end{equation}

           The quantity $\vec{\upsilon}_j(R,t)$ describe the current of probability
connected with the motion of $j-th$ particle, in general case
$\vec{\upsilon}_j(R,t)$ depend on coordinate of all particles of the system
$R$, where $R$ is the totality of $3N$ coordinate of $N$ particles
of the system $R=(\vec{r}_{1},...,\vec{r}_{N})$.

  The $S(R,t)$ value in the formula \ref{z4} represents the phase
of the wave function and as
the electron has spin, the wave function is now be expressed in the form
\begin{equation}\label{psi}
\psi(R,t)=a(R,t)e^{\frac{iS}{\hbar}}\varphi(R,t),
\end{equation}
   where $\varphi$, normalized such that $\varphi^+\varphi=1$, is the new spinor, defined in the local frame of reference with the origin at the point $\vec{r}$. The spinor gives the spin part of the wave function.

             We substituted the wave function in the definition of the basic hydrodynamical quantities. Using that the  velocity field $\vec{\upsilon}$ is the velocity of the local center of
mass and determined by equation

   \begin{equation}\label{vvv}
\vec{j}(\vec{r},t)=\rho(\vec{r},t)\vec{\upsilon}(\vec{r},t),
\end{equation}

the vorticity density field \ref{J2} and the  momentum current density tensor \ref{P} have the new form of

            \begin{equation}
\vec{\Omega}(\vec{r},t)=(\vec{\nabla}\times\vec{j})(\vec{r},t),           \end{equation}

\begin{equation}
\Re^{\alpha\beta}(\vec{r},t)=m\rho(\vec{r},t)\upsilon^{\alpha}(\vec{r},t)\upsilon^{\beta}(\vec{r},t)+\wp^{\alpha\beta}(\vec{r},t)+\end{equation}
\[ \qquad\qquad\qquad \qquad\qquad\qquad+\Lambda^{\alpha\beta}(\vec{r},t)+\Upsilon_s^{\alpha\beta}(\vec{r},t),
  \]
where
                     \begin{equation}
\wp^{\alpha\beta}(\vec{r},t)=\sum_{S}\int dR\sum^{N}_{j=1}\delta(\vec{r}-\vec{r}_{j})a^{2}m_ju^{\alpha}_ju_j^{\beta},
\end{equation}
is the well known kinetic pressure tensor.
Value $u^{\alpha}_{j}(\vec{r},R,t)$ is a
quantum equivalent of the thermal speed and $u^{\alpha}_{j}(\vec{r},R,t)=\upsilon^{\alpha}_{j}(R,t)-\upsilon(\vec{r},t)$.

                              The tensor $\Lambda^{\alpha\beta}$ is proportional to $\hbar^{2}$, has a purely quantum origin and can therefore be interpreted as an additional quantum pressure

                                \begin{equation}
\Lambda^{\alpha\beta}(\vec{r},t)=-\sum_{S}\int dR\sum^{N}_{j=1}\delta(\vec{r}-\vec{r}_{j})a^{2}\frac{\hbar^{2}}{2m_j}\frac{\partial^{2}\ln a}{\partial x_j^{\alpha}\partial x_j^{\beta}} \end{equation}

The quantum tensor \ref{TL} is the quantity which can be rewritten in  terms of concentration $\rho$ in the approximation of noninteracting particles, using the definition $\rho=\sum_{S}\int dR\sum^{N}_{j=1}\delta(\vec{r}-\vec{r}_{j})a^{2}(R,t)$     as
            \begin{equation}  \label{TL2}
\Lambda^{\alpha\beta}(\vec{r},t)=-\frac{\hbar^{2}}{4m}\rho (\vec{r},t)\partial^{\alpha}\partial^{\beta}(\ln\rho)(\vec{r},t)\end{equation}

using simple manipulation with the expression \ref{TL2} we may replace for
the large system of noninteracting particles, this tensor is
                  \begin{equation}  \label{TL}
\Lambda^{\alpha\beta}(\vec{r},t)=-\frac{\hbar^{2}}{4m}(\partial^{\alpha}\partial^{\beta}\rho(\vec{r},t)- \end{equation}                                                                             \[\qquad\qquad\qquad\qquad\qquad-\frac{1}{\rho(\vec{r},t)}\{\partial^{\alpha}\rho(\vec{r},t)\}\{\partial^{\beta}\rho(\vec{r},t)\})\]

It should be explained that the tensor  \ref{TL} arises as a consequence of the quantum  Madelung potential and can be  interpreted as an additional quantum pressure.

The tensor $\Upsilon_s^{\alpha\beta}$ appears in the theory as a result of representations rotating electrons as an assembly of bodies continuously distributed in space. In the context of quantum hydrodynamics the force due to a new spin stress inside the fluid takes the form of

\begin{equation} \label{SL} \Upsilon_s^{\alpha\beta}=-\frac{\hbar^2}{4m\mu^2}M_{\gamma}\partial^{\alpha}\partial^{\beta}(\frac{M^{\gamma}}{\rho})\end{equation}

This new force emerges from the inhomogeneity of spin distribution and must be considered in the equation of motion, being the order of $\hbar^2$.

         On the other hand, after the presentation of the wave function in the exponential form the tensor of the magnetic moment flux density  takes the form of

                       \begin{equation}  \label{Jj}
\Im^{\alpha\beta}_{M}(\vec{r},t)=M^{\alpha}\upsilon^{\beta}(\vec{r},t)+\gamma_s^{\alpha\beta}(\vec{r},t), \end{equation} where

\begin{equation}  \label{JM5} \gamma^{\alpha\beta}_s(\textbf{r},t)=-\sum_{S}\int
dR\sum^{N}_{j=1}\delta(\textbf{r}-\textbf{r}_{j})\frac{2\mu_j}{m_j\hbar}a^{2}(R,t)
\varepsilon^{\alpha\mu\nu}s_{j}^{\mu}\nabla_{j}^{\beta}s_{j}^{\nu}  \end{equation}

          In the context of quantum hydrodynamics we have the additional spin torque

                            \begin{equation}  \label{JM5}
\gamma^{\alpha\beta}_s(\textbf{r},t)=-\frac{\hbar}{2m\mu_j}
\varepsilon^{\alpha\gamma\lambda}M_{\gamma}(\textbf{r},t)\partial_{\beta}
(\frac{M^{\lambda}(\textbf{r},t)}{\rho(\textbf{r},t)}).
\end{equation}

\subsection{Energy evolution equation}

The energy density taking into account the Coulomb and Spin-Spin interactions is given by \cite{1}, \cite{2}

\begin{equation} \label{Se}
                                             \varepsilon(\textbf{r},t)=\int
                                             dR\sum^{N}_{j}\delta(\textbf{r}-\textbf{r}_{j})
                                             \frac{1}{4m_{j}}\{\psi^{+}_s\vec{D}^{2}_{j}\psi_s+(\vec{D}^{2}_{j}\psi_s)^{+}\psi_s\}(R,t)
                                  \end{equation}

                               \[  +\int dR\sum^{N}_{i\neq k}\delta(\textbf{r}-\textbf{r}_{j})
                               \frac{1}{2}\psi^{+}_s\{q_{j}q_{k}G_{jk}-\mu^{2}_{j}\sigma^{\alpha}_{j}
                               \sigma^{\beta}_{k}F^{\alpha\beta}_{jk}\}\psi_s(R,t).
                               \]

                  Differentiation of \ref{Se} with respect to time and application of the Schrodinger equation with
Hamiltonian \ref{H} leads to the energy balance equation

\begin{equation} \label{Se1}
           \frac{\partial}{\partial
           t}\varepsilon(\textbf{r},t)+\vec{\nabla}\vec{Q}
           (\textbf{r},t)=qj_{\alpha}(\textbf{r},t)E^{\alpha}(\textbf{r},t)
           \end{equation}

            \[
                                        \qquad\qquad\qquad\qquad+
                                        J^{\alpha\beta}_M(\textbf{r},t)\partial_{\beta}B^{\alpha}
                                        (\textbf{r},t)+A(\textbf{r},t),\]                          where $A(\textbf{r},t)$  - is the density of internal force and $\textbf{Q}(\textbf{r},t)$ - is the internal energy
                                      flux density.

         The internal energy
                                      flux density    is given by

                                     \begin{equation}\label{SQ} \textbf{Q}(\textbf{r},t)=\int
dR\sum^{N}_{j}\delta(\textbf{r}-\textbf{r}_{j})
\frac{1}{8m^2_{j}}\{\psi^{+}_s\textbf{D}_{j}\textbf{D}^{2}_{j}\psi_s+(\textbf{D}_{j}\textbf{D}^{2}_{j}\psi_s)^{+}\psi_s
\end{equation}

\[\qquad\qquad\qquad+\textbf{D}^{+}_{j}\psi^{+}_s\textbf{D}^{2}_{j}\psi_s +
(\textbf{D}^{2}_{j}\psi_s)^{+}\textbf{D}_{j}\psi_s\}(R,t)\]

\[+\int dR\sum^{N}_{j\neq k}\delta(\textbf{r}-\textbf{r}_{j})\frac{1}{4m_{j}}
\{\psi^{+}_s(R,t)(q_{j}q_{k}G_{jk}\]

           \[ \qquad\qquad\qquad-\mu^{2}_{j}\sigma^{\alpha}_{j}\sigma^{\beta}_{k}F^{\alpha\beta}_{jk})\textbf{D}_{j}\psi_s(R,t)+k.c.\} \]

        The force density of internal forces  in (\ref{Se1}) consists of Coulomb force density $A_{cl}(\textbf{r},t)$ and Spin-Spin force density $A_{s-s}(\textbf{r},t)$

                     \begin{equation}\label{A}
    A_{s-s}(\textbf{r},t)=\int dR\sum^{N}_{i\neq k}\delta(\textbf{r}-\textbf{r}_{j})\frac{1}{4}(\nabla_{\gamma}F^{\alpha\beta}_{jk})\times
                                                                     \end{equation}

     \[\times\{\frac{1}{m_{j}}
     \{(\hat{\sigma}_{j}^{\alpha}\hat{\sigma}_{k}^{\beta}\hat{D}^{\gamma}_{j}\psi)^{+}\psi
     +\psi^{+}(\hat{\sigma}_{j}^{\alpha}\hat{\sigma}_{k}^{\beta}\hat{D}^{\gamma}_{j}\psi)\}(R,t)
     \]

     \[+\frac{1}{m_{k}}\{(\hat{\sigma}_{j}^{\alpha}\hat{\sigma}_{k}^{\beta}\hat{D}^{\gamma}_{k}\psi)^{+}\psi
     +\psi^{+}(\hat{\sigma}_{j}^{\alpha}\hat{\sigma}_{k}^{\beta}\hat{D}^{\gamma}_{k}\psi)\}(R,t)\}
                                                                     \]                                                                   

                                     and 
                                     
                                     \[ A_{cl}(\textbf{r},t)=
-\int dR\sum^{N}_{i\neq
k}\delta(\textbf{r}-\textbf{r}_{j})\frac{1}{4}(q_{j}q_{k}\nabla_{\alpha}G_{jk})\times
\]

                                                            \[ \times\{\frac{1}{m_{j}}\{\hat{D}^{+\alpha}_{j}\psi^{+}(R,t)\psi(R,t)+\psi^{+}(R,t)\hat{D}^{\alpha}_{j}\psi(R,t)
                                                            \]

                                                            \[+\frac{1}{m_{k}}\{\hat{D}^{+\alpha}_{k}\psi^{+}(R,t)\psi(R,t)+\psi^{+}(R,t)\hat{D}^{\alpha}_{k}\psi(R,t)\}
                                                            \}
                                                            \]

                           Using
the fact that the velocity field is the velocity of the local center of mass and is determined by  \ref{vvv} and using the definitions  \ref{SL} and \ref{JM5} the energy evolution equation reads

                     \begin{widetext}
     \begin{equation}   \label{Se3}
 \rho(\frac{\partial}{\partial
 t}+\textbf{v}\nabla)\epsilon(\textbf{r},t)+\vec{\nabla}\vec{q}(\textbf{r},t)+\frac{\hbar^{2}}{4m\rho(\textbf{r},t)}\{\partial_{\alpha}\rho(\textbf{r},t)\}\{\partial^{\beta}\rho(\textbf{r},t)\}\partial_{\beta}\upsilon^{\alpha}(\textbf{r},t)
                                                                                                                                                                       +p^{\alpha\beta}(\textbf{r},t)\partial_{\beta}\upsilon_{\alpha}(\textbf{r},t)\end{equation}

 \[  \qquad
 -\frac{\hbar^{2}}{4m}\partial_{\alpha}\partial^{\beta}\rho(\textbf{r},t)\partial_{\beta}\upsilon^{\alpha}(\textbf{r},t)-\frac{\hbar^2}{4m\mu^2}M_{\gamma}(\textbf{r},t)\partial_{\alpha}\partial^{\beta}(\frac{M^{\gamma}(\textbf{r},t)}{\rho(\textbf{r},t)})\partial_{\beta}\upsilon^{\alpha}(\textbf{r},t)
\]                                                                                                                                                                                                                                                                                                                                                                    \[\qquad\qquad\qquad\qquad=-\frac{\hbar}{2m\mu}\varepsilon^{\alpha\mu\nu}M_{\mu}(\textbf{r},t)\partial_{\beta}(\frac{M^{\nu}(\textbf{r},t)}{\rho(\textbf{r},t)})\nabla_{\beta}B^{\alpha}_{ext}(\textbf{r},t) +\aleph(\textbf{r},t). \]
                \end{widetext}

               Let us discuss the physical significance of the terms on the right-hand side of the system of MQHD
equation  \ref{Se3}. The third  and fifth terms on the right-hand side in Eq.  \ref{Se3} describe a quantum force produced by density fluctuations, which has its origin in the
so-called Madelung potential. The forth term represents   the well
known pressure tensor influence. The sixth term characterizes the energy density generation by the spin stress and the seventh term on the left-hand side  of \ref{Se3} describes the magnetic moment density torque influence. 

     The relative energy density  the of spinning fermions in an external electromagnetic field takes the form

                     \begin{widetext}
               \begin{equation} \label{Se2}
  \rho\epsilon(\textbf{r},t)=\int
  dR\sum^{N}_{j}\delta(\textbf{r}-\textbf{r}_{j})a^2(R,t)(\frac{m_j\textbf{u}^2_j}{2}-
  \frac{\hbar^2}{2m_j}\frac{\triangle_ja}{a}+\frac{1}{2m_j}|\nabla_{\alpha}s^{\alpha}_j|^2)
  \end{equation}
  \[  \qquad\qquad+ \frac{q^2}{2}\int
  d\textbf{r}^{'}G(\textbf{r},\textbf{r}^{'})\rho_2(\textbf{r},\textbf{r}^{'},t)- \frac{1}{2}\int
  d\textbf{r}^{'}F^{\alpha\beta}(\textbf{r},\textbf{r}^{'})M^{\alpha\beta}(\textbf{r},\textbf{r}^{'},t)
  \]                          \end{widetext}                                                                           

    The first term on the right-hand side of the expression  \ref{Se2}  describes the quantum equivalent of the thermal speed contribution, the second term   characterizes the quantum Madelung potential   contribution and the third term presents the internal spin potential influence. The forth and fifth terms describe a force
field that represents interactions between particles, namely the Coulomb interaction of charges and
spin–spin interactions.

Note that to simplify the problem we consider that the thermal spin-interactions are neglected and   microscopic spin  $s^{\alpha}_j=s^{\alpha}$ is equal to macroscopic average $s^{\alpha}$.
              Taken in the approximation
of self-consistent field, from \ref{jjj}, \ref{magnit}  we have the set of MQHD equation for the $electrons$ and $positrons$ (p=e, i): continuity
equation, momentum balance equation,   magnetic moment density equation take the form

                               \begin{equation} \label{nnn}
\partial_{t}\rho_p+\vec{\nabla}(\rho_p\vec{\upsilon}_p)=0,
\end{equation}

                       \begin{widetext}
            \begin{equation} \label{jjjjj}
m_p\rho_p(\partial_{t}+\upsilon^{\beta}_p\partial_{\beta})\vec{\upsilon}_p=q_p\rho_p\vec{E}_{ext}+\frac{1}{c}\vec{j}_{pe}\times\vec{B}_{ext}-\vec{\nabla}\wp_p\end{equation}   \[+\frac{\hbar^2}{2m_p}\rho_p\vec{\nabla}(\frac{\triangle\sqrt{\rho_p}}{\sqrt{\rho_p}}) +M_{p\beta}\vec{\nabla}B^{\beta}_{ext}+\frac{\hbar^2}{4m\mu^2}\partial_{\beta}\{M_p^{\gamma}\vec{\nabla}\partial^{\beta}(\frac{M_p^{\gamma}}{\rho_p})\}        \]
\[-\rho_p\vec{\nabla}\int d\vec{r}^{'}q^2_pG(\vec{r},\vec{r}^{'})\rho_p(\vec{r}^{'},t)+M_{p\gamma}\vec{\nabla}\int d\vec{r}^{'}F^{\gamma\delta}(\vec{r},\vec{r}^{'})M^{\delta}_p(\vec{r}^{'},t),
  \]                         \end{widetext}
  
           \begin{widetext}
 \begin{equation} \label{Mm}
(\partial_{t}+\upsilon^{\beta}_p\partial_{\beta})\vec{M}_p=\frac{2\mu_p}{\hbar}\vec{M}_{p}\times\vec{B}_{ext} +\frac{\hbar}{2m_p\mu_p}\partial_{k}\{\vec{M}_{p}\times\partial^{k}(\frac{\vec{M}_p}{\rho_p})\}+\frac{2\mu_p}{\hbar}\epsilon^{\alpha\beta\gamma}M^{\beta}_p\int d\vec{r}^{'}F^{\gamma\delta}(\vec{r},\vec{r}^{'})M^{\delta}_p(\vec{r}^{'},t),
\end{equation}
                   \end{widetext}

                                                        Let's discuss the physical significance of terms on the
right side of the system of MQHD equations obtained above \ref{nnn} - \ref{Mm}.    The first and second terms in Eq. \ref{jjjjj}describe the well
known interaction with the external electromagnetic field, where the first term represents the effect
of the external electric field on the charge density and the second term is the Lorentz force field.
The fourth term is a quantum force produced by density fluctuations, which has its origin in the
so-called Madelung potential. The fifth term appears in the equation of motion \ref{jjjjj} through the
magnetization energy and depends on the spin or magnetic moment density of particles. The sixth
term represents the self-force or magnetic moment density stress inside the electron or positron fluid.
This spin self-force appears even in the absence of the electromagnetic fields and arises from the
inhomogeneity of the magnetic moment density distribution. Other terms in \ref{jjjjj} describe a force field that represents
interactions between particles, namely the $Coulomb$
interaction of charges and $Spin-Spin$ interactions.

The second term in the equation of magnetic moment density
motion \ref{Mm}
represents the effect additional  $magnetic$ $moment$ $density$ $torque$ on the magnetic moment density evolution and tends to align spins parallel.  It's important  that the second term has a similar form respectively to the the contribution of  exchange interaction in ferromagnetic media for isotropic cubic ferromagnetic.

Using the definition \ref{J2} and the Madelung decomposition \ref{z4} with the  momentum balance  dynamical equation \ref{jjjjj}
 the  hydrodynamics  classical vorticity dynamical equation  $\vec{\omega}_p=\vec{\nabla}\times\vec{\upsilon}_p$

          \begin{equation} \label{v}
\partial_{t}\vec{\omega}_p=\vec{\nabla}\times(\vec{\upsilon}_p\times\vec{\omega}_p)-\vec{\nabla}(\frac{1}{\rho_p})\times\vec{\nabla}\wp_p+\frac{1}{m_p}\vec{\nabla}(\frac{M_{pk}}{\rho_p})\times\vec{\nabla}B^k_{ext} \end{equation}
\[+\frac{1}{cm_p}\vec{\nabla}\times(\frac{1}{\rho_p}\vec{j}_{pe}\times\vec{B}_{})-\frac{q^2_p}{m_p}\vec{\nabla}\times\vec{\nabla}\int d\vec{r}^{'}G(\vec{r},\vec{r}^{'})\rho_p(\vec{r}^{'},t)\]
\[   +\frac{\hbar^2}{4m_p^2\mu^2_p}\vec{\nabla}(\frac{M_p^{\nu}}{\rho_p})\times\vec{\nabla}\{\frac{1}{\rho_p}\nabla_k(\rho_p\nabla^k\{\frac{M^{\nu}_p}{\rho_p})\} \] \[+ \frac{1}{m_p}\vec{\nabla}(\frac{M_{p\gamma}}{\rho_p})\times\vec{\nabla}\int d\vec{r}^{'}F^{\gamma\delta}(\vec{r},\vec{r}^{'})M^{\delta}_p(\vec{r}^{'},t).
\]

The vorticity  evolution equation \ref{v} shows the different physical
factors associated with the generation of vorticity. The second term on the
right side of \ref{v}  is proportional to the gas
pressure and is responsible for the hydrodynamic baroclinic vorticity generation of the classical
vortex field. The third term represents the magnetic baroclinic vorticity and is associated with the
anisotropic magnetic pressure effect. The fourth term contains information about the vorticity generated
by the magnetic tension. The sixth term is associated with the magnetic vorticity generation,
even in the absence of the magnetic field. The sixth and seventh terms   characterize the effect of  $Coulumb$ and $Spin-Spin$  interactions in the vorticity evolution. Equation \ref{v}  contains the normal electron or positron
current density $\vec{j}_{ep}=q_p\rho_p\vec{\upsilon}_p$, and  the  magnetic moment density $\vec{M}_p=\rho_p\vec{\mu}_p$. The vorticity  evolution equation \ref{v}  is a generalization of classical vorticity equation which had been presented in works \cite{12},  \cite{13}, \cite{22} and \cite{24}. At first, Eq. \ref{v} combines the
erstwhile generalized classical vorticity, but in contrast to \cite{13} and \cite{24} contains the information about interactions inside the quantum vortical fluid and have been derived using the MQHD method.

Note, that for a 3D system of particles the momentum
balance equation \ref{jjjjj}, the magnetic density equation \ref{Mm} and the vorticity evolution equation \ref{v} may be written down in terms of
magnetic intensity of the field that is created by charges $q_p$
and spins $\vec{s}_p$ of the particle system

                     \begin{equation} \label{j2}
m_p(\partial_{t}+\upsilon^{\beta}_p\partial_{\beta})\vec{\upsilon}_p=q_p\vec{E}+\frac{q_p}{c}\vec{\upsilon}_{pe}\times\vec{B}-\frac{\vec{\nabla}\wp_p}{\rho_p}+\frac{\hbar^2}{2m_p}\vec{\nabla}(\frac{\triangle\sqrt{\rho_p}}{\sqrt{\rho_p}}) \end{equation}
\[\qquad\qquad\qquad\qquad\qquad\qquad+\frac{2\mu_p}{\hbar}s_{p\beta}\vec{\nabla}B^{\beta}_{eff},\]

               \begin{equation} \label{Mm2}
(\partial_{t}+\upsilon^{\beta}_p\partial_{\beta})\vec{s}_p=\frac{2\mu_p}{\hbar}\vec{s}_{p}\times\vec{B}_{eff} ,
\end{equation}

and
 \begin{equation} \label{v2}
\partial_{t}\vec{\tilde{\omega}}_p=\vec{\nabla}\times(\vec{\upsilon}_p\times\vec{\tilde{\omega}}_p)-\vec{\nabla}(\frac{1}{\rho_p})\times\vec{\nabla}\wp_p+\frac{2\mu_p}{\hbar m_p}\vec{\nabla}s_k\times\vec{\nabla}B^k_{eff}   \end{equation}
where $\vec{\tilde{\omega}}_p=\vec{\omega}_p+\frac{q_p}{m_pc}\vec{B}$ - is the generalized vorticity and the effective magnetic field $\vec{B}_{eff}=\vec{B}+\vec{B}_{in}$ includes the total magnetic field and  internal magnetic field $\vec{B}_{in}$

      \begin{equation} \label{B2}
  \vec{B}_{in}=\frac{c}{q_p\rho_p}\nabla_k(\rho_p\nabla^k\vec{s}_{p})
\end{equation}

The total magnetic field $\vec{B}$  consists of the field generated by the charge and the field generated by the spins.   Amp`ere's law        including the magnetization spin current   $j_m=2\mu/\hbar\vec{\nabla}\times(\rho\vec{s})$   takes the form of

                             \begin{equation} \label{M1}
\vec{\nabla}\times B=\frac{4\pi}{c}\sum_{p}\vec{j}_{pe}+\frac{8\pi\mu}{\hbar}\sum_{p}\vec{\nabla}\rho_p\times\vec{s}_p+\frac{8\pi\mu}{\hbar}\sum_{p}\rho_p\vec{\nabla}\times\vec{s}_p, \end{equation}

We must note that the spin stress term have notably interesting nature,  exist even in absence of magnetic field, have only quantum foundation and arising out from the spin part of the wave function. Equation \ref{v2} was rewritten  by separating the magnetic
and non-magnetic terms \cite{8}. effective magnetic field and the last three terms produced by the magnetization vortex generation.
Lets to rewrite the spin-vortex evolution equation from \ref{vs}, using the vector $\vec{\Xi}=\vec{\nabla}\times\vec{s}$

\begin{equation} \label{vs2}
\partial_{t}\vec{\Xi}=\vec{\nabla}\times(\vec{\upsilon}\times\vec{\Xi})+\frac{2\mu}{\hbar}\vec{\nabla}\times(\vec{s}\times\vec{B}_{eff}). \end{equation}

This is new equation was obtained with the method of magneto quantum hydrodynamics (MQHD) for the study of the quantum
evolution of a system of spinning fermions.

\subsection{The Whistler Mode Turbulence in Magnetized Plasmas}
The nonlinear turbulent processes associated with electromagnetic waves in spinning plasmas have
attracted interest. Nonlinear whistler mode turbulence has been studied in a magnetized plasma \cite{25} - \cite{28}. The authors  had focused on low-frequency (in comparison with the electron gyro-frequency) nonlinearly interacting electron whistlers and nonlinearly interacting Hall-magnetohydrodynamic (H-MHD) fluctuations in \cite{25}. In this section we investigate the electron whistler wave properties
based on extended 2D magnetohydrodynamic equations. However, we understand that the electron
spin effect on the whistler wave dispersion typically requires a strong external magnetic field.

Two-dimensional turbulence has been studied in a magnetized plasma involving incompressible
electrons and immobile ions. We consider that the electrons carry currents, while the immobile ions
provide a neutralizing background to a quasi-neutral spinning plasma. Using the fact that the electron fluid velocity is associated not only with the rotational magnetic field but also with the magnetization spin current $j_M=2\rho_0\mu_e\vec{\nabla}\times \vec{s}/\hbar$ which is determined by the spin vector $\vec{s}$, we have from the Amp`ere's law

                           \begin{equation} \label{VB}
\vec{\upsilon}_e=-\frac{c}{4\pi\rho_0e}\vec{\nabla}\times \vec{B }-\frac{g}{2m_e}\vec{\nabla}\times \vec{s}
\end{equation}
where $\mu_e=-ge\hbar/4m_ec$, $m_e$ - is the electron mass, $\rho_0$ - is the electron density. We take into account that the electron density is constant and the electron continuity equation \ref{nnn} shows a divergence-less electron fluid velocity $\vec{\nabla}\vec{\upsilon}_e=0.$

All physical
quantity is presented in the form of sum of equilibrium part
and small perturbations  $f=f_0+f_1$
\begin{align}
\vec{B}(\vec{r},t)=B_{0}\vec{y}+\vec{B}_1(\vec{r},t)+... \qquad   \vec{s}_e(\vec{r},t)=s_{0}\vec{y}+\vec{s}_1(\vec{r},t)
\end{align}
\[
       \rho_e(\vec{r},t)=\rho_{0}.., \qquad  \vec{\upsilon}_e(\vec{r},t)=\vec{\upsilon}_1(\vec{r},t)+.., \] \[ \vec{\omega}_e(\vec{r},t)=\vec{\omega}_1(\vec{r},t)+..,\]
 where $B_0$ - is the external uniform magnetic field directed along the axis $y$,  $s_{0}$  - is the unperturbed   spin vector.

In this case if we assume that linear excitations $f_1$ are
proportional to $\exp(-i\omega t+i\vec{k}\vec{x})$, where $\omega$ - is the wave frequency and $k^2=k_x^2+k_y^2$.  The three-dimensional equation \ref{v2} closed by \ref{Mm2}  transformed into two dimensional by the regarding variation in the $\vec{z}$-direction as ignorable or $\partial/\partial z=0$ and used the separation for the total magnetic field into two scalar variables $\vec{B}_1=\vec{z}\times\vec{\nabla}\psi+b\vec{z}$ \cite{25}.

We will assume propagation of the waves along an external magnetic field $B_{0}$ or $k=k_y.$ A linearized set of equations    \ref{Mm2} and \ref{v2} in this case  gives us the dispersion equation
                    \begin{widetext}
         \begin{equation} \label{D1}
(1+k^2)\omega^3_k - (1-\omega_{\mu})k^2\omega^2_k + ((\omega_{\mu}\tilde{\omega}_{g}-\tilde{\omega}^2_{g}-\omega_{\mu})k^2-\tilde{\omega}^2_{g})\omega_k+\tilde{\omega}^2_{g}k^2-\tilde{\omega}_{g}\omega_{\mu}k^2=0, 
\end{equation}     \end{widetext}
where the length and time scales are normalized respectively $d_e=c/\omega_{pe}$ and $\omega_c=eB_0/m_ec$, $d_e$ - is the electron skin depth or inertial length scale, $\omega^2_{pe}=4\pi e^2\rho_0/m_e$ - is the electron
plasma frequency  $\omega_c$ - is the electron  cyclotron frequency   and $c$ - is the speed of light. The other physical quantities are normalized as \[\omega_k\rightarrow\frac{\omega_k}{\omega_c},\qquad k\rightarrow kd_e \qquad \tilde{\omega}_g\rightarrow\frac{\tilde{\omega}_g}{\omega_c}, \qquad \omega_\mu\rightarrow\frac{\omega_\mu}{\omega_c}\]

\[\tilde{\omega}_g=g\omega_c/2+k^2|s_0|/m_e\] - is   the spin-precession frequency which includes the internal magnetic field influence and $\omega_\mu=g^2|s_0|/4m_ed^2_e$ - is a frequency that involves a spin correction due to the plasma magnetization
current and appears even in the absence of the external magnetic field $B_0$, $\hbar$ - is the reduced Planck
constant. We use that an unperturbed spin state $s_0=-\hbar/2\tanh(\mu_BB_0/k_BT_e)$ antiparallel to the background magnetic field. This function appears as the
solution of the spin evolution equation for spin quantum plasmas where the spin inertia and the spin thermal
coupling terms are neglected \cite{4}, \cite{29}. The temperature $T_e$ is  the Fermi
electron temperature $T_F=\hbar^2(3\pi^2\rho_0)^{2/3}/2m_ek_B,$ where $k_B$ is the Boltzmann constant. A situation magnetization effects might be important in a regime of very strong magnetic field in which the external field strength
approaches or exceeds the quantum critical magnetic field $B_0\sim4.4138\times10^{13}G$ and highly dense plasmas $\rho_0\sim10^{30}1/sm^{3}$. But it can be assumed that   the internal magnetic field inside the fluid which is dependent on the gradient of the spin distribution \ref{B2}, can tend to align neighboring spins parallel.

The effect of the frequency that involves the spin correction due to the plasma magnetization current
is small  $\omega_\mu<\omega_c$  the cubic
 expression \ref{D1} may be expanded to yield formulae
$\omega_k$ in the following form

              \begin{equation} \label{D2}
\omega_1=\frac{k^2}{1+k^2}(1-\omega_\mu)+\frac{\omega_\mu(\tilde{\omega}_g-1)k^2}{(\tilde{\omega}_g+\omega_\mu-1)k^2+\tilde{\omega}_g},
\end{equation}
             and
\begin{equation} \label{D3}
\omega_2=\tilde{\omega}_g+\frac{\omega_\mu(\tilde{\omega}_g-1)k^2}{(\tilde{\omega}_g+\omega_\mu-1)k^2+\tilde{\omega}_g},
\end{equation}

The relation \ref{D2}   expresses the dispersion of low-frequency whistler waves
in the spinning quantum plasma using the model based on the 2D electromagnetic turbulence
equation \ref{v2}.
The solution \ref{D3} expresses the dispersion of waves that
emerge as a result of spin dynamics.  The spectrum is divided by the electron inertial skin depth into two regions, short scale $kd_e > 1$, $ \omega_k\sim 1$  and long scale region $kd_e < 1$, $ \omega_k\sim k^2.$

                     \section{Conclusions}

In this paper we analyzed $vorticity$  excitations caused by the
magnetic moment density dynamics in systems of charged  $ 1/2-spin$ particles.   MQHD equations are a consequence of MPSE in which particles'
interaction is directly taken into account. In our work
we consider the $Coulomb$ and $Spin-Spin$
interactions. The system of MQHD equations we have constructed comprises equations of continuity,
of the momentum balance, of the energy evolution equation, of the magnetic moment density evolution, and of the vorticity density
dynamics. In our studies of wave processes we have used a self-consistent field approximation of the
MQHD equations.

The equations we are interested in, determining the system dynamics, are the hydrodynamic equations for the spinning plasma. This equations (\ref{jjjjj} and \ref{Mm})
have an additional quantum contribution proportional to $\hbar^2$ and spin corrections, additional $Magnetic$ $Moment$ $Stress$  and $Magnetic$ $Moment$  $Torque$ which have been derived in the absence of (thermal) fluctuation of the spin about the macroscopic average. But in such a situation (thermal) effects on the spin might be important. The main objective of this paper was
to construct an appropriate a new generalized  vorticity equation \ref{v2} for  spin quantum plasmas that contains the magnetic, non-magnetic terms and  the spin dependent forces being non
potential. The turbulent processes in plasmas had been investigated using the vorticity equation   \cite{12}, \cite{13}, \cite{14}. We had derived the vortex dynamic formulation of spinning non - relativistic quantum plasma, using the method of magneto quantum hydrodynamics (MQHD). We had generalized the classical vorticity equation for a spinning quantum fluid plasma and derived the vorticity equation \ref{v} in which particles'
interactions ( $Coulomb$ and $Spin-Spin$) is directly taken into account.    Important that the
quantum Madelung potential  do not contribute
to the vorticity evolution.

Using MQHD equations we analyzed elementary excitations
in various physical systems in a linear approximation. We had studied the  influence of the intrinsic spin of electrons in the nonlinear whistler mode turbulence. Dispersion branches characterize a  new waves, one of which propagates below
the electron cyclotron frequency \ref{D2}, one above the spin-precession frequency due to the spin perturbations \ref{D3}. This result had been derived for the incompressible electrons in the model based on the two-dimensional vorticity equation.  The spin effects are seen to be substantial in the very strong magnetic
field,   dense plasmas  and the graphical representation of the waves is similar to that found in \cite{}

The investigation of this approach leads to interesting spin effects dense quantum plasmas  in compact astrophysical objects,  plasmas in semiconductors and micro-mechanical
systems,  in quantum x-ray free-electron lasers.

 \appendix
\section{}

The  terms represented $Coulomb$ and $Spin-Spin$
interactions in Eq. \ref{jjjjj}, \ref{Mm} and \ref{v} leads to the appearance of the self-consistent electric field $E_{int}$ and the self-consistent  magnetic field $B_{spin}$

            \begin{subequations}
\begin{eqnarray}
  \label{E}
\vec{\nabla}\vec{E}_{int}=4\pi q\rho,  \qquad  \vec{\nabla}\times\vec{E}_{int}=0,
\end{eqnarray}
\end{subequations}
            \[   \vec{\nabla}\vec{B}_{spin}=0,  \qquad  \vec{\nabla}\times\vec{B}_{spin}= 4\pi\vec{\nabla}\times\vec{M}.
            \]

         The two-particle hydrodynamics functions can be produced using the self-consistent field method.    Two-particle functions \ref{F} and \ref{F2} have the ground expressions  \cite{1}, \cite{2}

                               \begin{equation}   \rho_2(\vec{r},\vec{r}^{'},t)=\rho(\vec{r},t)\rho(\vec{r}^{'},t) +\varrho(\vec{r},\vec{r}^{'},t),   \end{equation}
                                \begin{equation}   M^{\alpha}_2(\vec{r},\vec{r}^{'},t)=M^{\alpha}(\vec{r},t)M^{\alpha}(\vec{r}^{'},t) +\chi^{\alpha}(\vec{r},\vec{r}^{'},t),   \end{equation}
                                where $\varrho(\vec{r},\vec{r}^{'},t)$  and  $\chi^{\alpha}(\vec{r},\vec{r}^{'},t)$ - are the the correlation functions. We must note that  the two-particle functions are the functionals of the wave function $\varphi(R,t)$.

               \end{document}